\documentclass[prl,twocolumn,epsf]{revtex4}
\usepackage{graphicx}
\usepackage{dcolumn}
\usepackage{bm}
\usepackage{hyperref}
\renewcommand{\vec}[1]{{\mathbf{#1}}}
\newcommand{\beq}{\begin{eqnarray}}
\newcommand{\eeq}{\end{eqnarray}}

\begin{document}

\title{Hidden Charge 2e Boson in Doped Mott
Insulators: Field Theory of Mottness}

\author{Robert G. Leigh}
\author{Philip Phillips}
\author{Ting-Pong Choy}
\affiliation{Department of Physics,
University of Illinois
1110 W. Green Street, Urbana, IL 61801, U.S.A.}

\date{\today}

\begin{abstract}
We construct the low energy theory of a doped Mott insulator, such as
the high-temperature superconductors, by explicitly integrating over the
degrees of freedom far away from the chemical potential. For either hole
or electron doping, a
charge 2e bosonic field emerges at low energy. The charge 2e boson mediates dynamical spectral weight transfer
across the Mott gap and creates a new charge $e$ excitation by binding a hole. The result is a bifurcation
of the electron dispersion below the chemical potential as observed
recently in angle-resolved photoemission on Pb-doped Bi$_2$Sr$_2$CaCu$_2$O$_{8+\delta}$ (Pb2212).
\end{abstract}

\maketitle


Two problems beset the construction of a proper low-energy theory
(explicit integration of the high energy scale) of doped Mott
insulators.  First, the high energy degrees of freedom are neither
fermionic nor bosonic.  To illustrate, in  a Mott insulator, the
chemical potential lies in a charge gap between two bands that represent
electron motion on empty (lower Hubbard band, LHB for short) and singly
occupied sites (upper Hubbard band, hereafter UHB).  Since the latter
involves double occupancy, the gap between the bands is set by the
on-site repulsion energy, $U$. Nonetheless, both double occupancy and
double holes represent high energy excitations in the half-filled
insulating state as each is equally far from the chemical potential.  As
neither of these is fermionic, standard fermionic path integral
procedures are of no use.

Second, unlike the static bands in band insulators, the UHB
and LHB are not rigid, thereby permitting spectral weight transfer.
When $x$ holes are placed in a Mott insulator, at least $2x$\cite{sawatzky} single particle
addition states are created just above the chemical
potential.  The deviation from $x$, as would be the case
in a band insulator, is intrinsic to the strong correlations that
mediate the Mott insulating state in a half-filled band, thereby
distinguishing Mottness from ordering. The states in excess of $x$ arise
from two distinct effects. Each hole reduces the number of ways of creating
a doubly occupied site by one, thereby reducing the spectral weight at high energy.  As the $x$ empty sites can be occupied by either
spin up or spin down electrons, the $2x$ sum rule
is exact\cite{sawatzky} in the atomic limit.  Further, in the presence of
hybridization (with matrix element $t$), virtual excitations between the
LHB and UHB increase the loss of spectral weight at high energy thereby
leading to a faster than $2x$ growth\cite{sawatzky,harris,eskes} of the
low-energy spectral weight, a phenomenon
confirmed\cite{cooper,uchida1,chenbatlogg} widely in the high-temperature copper-oxide
superconductors.

Because some of the low-energy degrees of freedom of doped Mott insulators derive from the high
energy scale, low-energy descriptions must either C1)
abandon Fermi statistics or C2) generate new degrees of freedom\cite{sawatzky}
which ultimately leads to electron number non-conservation.
Current proposals for the low-energy physics of doped Mott insulators
are based either on perturbation theory\cite{Girvin} followed by
projecting out the high-energy sector or slaved\cite{slaveboson}
operators designed to exclude double occupancy.  As projection is not
integration, neither permits an explicit integration of
the high energy scale, and both miss relevant physical aspects.

We show that exact integration of the
high energy scale results in a low-energy theory that possesses
a charge $2e$ bosonic mode. Such an excitation might have been anticipated in light of the mixing between high and low energy multiply charged states\cite{slavery}. Our theory is an explicit example of C2
as the conserved charge involves both the boson and electron number. Note that the {\it emergence} of new degrees of freedom in a low energy theory, not directly built out of elementary excitations, is not without precedent. Indeed we believe that there are useful lessons to draw for Mott insulators from analogies with confining theories or other strongly coupled theories. A simple theoretical model which bears some resemblance to the theory that we develop below is the non-linear $\sigma$-model (NL$\sigma$-M), in which an initially non-dynamical field develops correlations and in fact determines the phase structure of the theory. 

While our starting point is the one-band Hubbard model,
\beq\label{hubbham}
H_{\rm Hubb}=-t\sum_{i,j,\sigma} g_{ij} c^\dagger_{i,\sigma}c_{j,\sigma}
+U\sum_{i,\sigma} c^\dagger_{i,\uparrow}c^\dagger_{i,\downarrow}c_{i,\downarrow}c_{i,\uparrow}
\eeq
our scheme is completely general and is applicable to the n-band case as
well. Here $i,j$ label lattice sites, $g_{ij}$ is equal to one iff $i,j$
are nearest neighbours and $c_{i\sigma}$ annihilates an electron with
spin $\sigma$ on lattice site $i$.   The Hilbert space of this model is
a product of Fock spaces, $\otimes_i \left({\cal F}_\uparrow\otimes
{\cal F}_\downarrow\right)$. We are concerned in the limit when the
Hubbard bands are well-separated,  $U\gg t$.  Given that the chemical
potential lies in the gap between such well separated bands at
half-filling, the high energy degree of freedom is ambiguous at
half-filling. Both double occupancy (UHB) and double holes (LHB) are
equally costly.  Doping removes this ambiguity. Hole-doping jumps the
chemical potential to the top of the LHB thereby defining double
occupancy to be the high energy scale. For electron doping, the chemical
potential lies at the bottom of the upper Hubbard band and it is the
physics associated with double holes in the lower Hubbard band that should
be coarse-grained.  A generalized particle-hole transformation
relates the two theories.

\begin{figure}
\centering
\includegraphics[width=8.8cm]{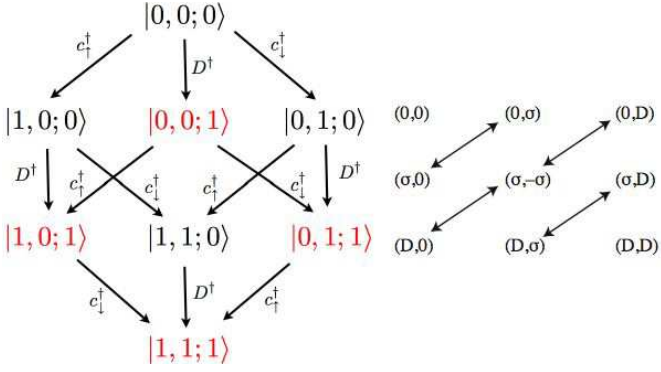}
\caption{Extended Hilbert space (left) which allows an explicit
integration of the high energy scale. Hopping processes (right)
included in the Lagrangian.}
\label{hilb}
\end{figure}

\noindent {\bf Hole Doping:} The basic idea of our construction is to rewrite the Hubbard model in such a way as to isolate the high energy degrees of freedom so that they can be simply integrated out. We do this by first introducing a new oscillator that represents the degrees of freedom at high energy and 
including a constraint which ensures that the extended theory is {\it equivalent} to the Hubbard model. If we simply solve this constraint we return to the description (1) of the Hubbard model, while if instead, we integrate out the high energy degrees of freedom, we will obtain the low energy effective theory. To this end, we extend the Hilbert space $\otimes_i \left({\cal F}_\uparrow\otimes {\cal F}_\downarrow\otimes
{\cal F}_D\right)$. We associate $D^\dagger$ with the creation of double-occupation.   In order to limit the Hilbert
space to single occupation in the $D$ sector, we will take $D$ to be
fermionic.  Integrating over the high-energy scale will be
accomplished by integrating over $D$. In particular, we will formulate a Hamiltonian for the extended theory in such a way that if we were to solve the constraint, precisely the Hubbard Hamiltonian (1) would be recovered. The action of the standard electron creation operator,
$c^\dagger_{i\sigma}$ and the new fermionic operator, $D^\dagger$,
to create the allowed states on a single site are shown in Fig.
\ref{hilb}.
We now formulate a Lagrangian
\beq\label{LE}
L&&=\int d^2\eta\left[\bar{\eta}\eta\sum_{i\sigma}(1- n_{i\bar\sigma})
 c^\dagger_{i\sigma}\dot c_{i\sigma}
 +\sum_i D_i^\dagger\dot D_i\right.\nonumber\\
&&+U\sum_j D^\dagger_jD_j
- t\sum_{i,j,\sigma}g_{ij}\left[ C_{ij\sigma}c^\dagger_{i,\sigma}c_{j,\sigma}
+D_i^\dagger c^\dagger_{j,\sigma}c_{i,\sigma}D_j\right.\nonumber\\
&&+\left.\left.(D_j^\dagger \eta c_{i,\sigma}V_\sigma c_{j,\bar\sigma}+h.c.)
\right]+H_{\rm con}\right],
\eeq
in the extended Hilbert space in such a way as to include the hopping
terms (see right panel in Fig. \ref{hilb}) present in the Hubbard
model (where we have replaced doubly occupied sites by $D$-occupation). Here, $\eta$ is a formal complex Grassmann constant which we have
inserted in order to keep track of statistics, and $d^2\eta$ denotes Grassmann integration.
The parameter $V_\sigma$ has values $V_\uparrow =1$,
$V_\downarrow=-1$, and simply ensures that $D$ couples to the spin
singlet. The operator $C_{ij\sigma}$ is of the form $C_{ij\sigma}\equiv
\bar\eta\eta\alpha_{ij\sigma}\equiv
\bar\eta\eta(1-n_{i,\bar\sigma})(1-n_{j,\bar\sigma})$ with number operators
$n_{i,\sigma}\equiv c^\dagger_{i,\sigma}c_{i,\sigma}$. Note that the
dynamical terms that appear in the Lagrangian are non-traditional
because the dynamics with the $c_{i\sigma}$ operators must exclude those
sites which contain the occupancy
$c^\dagger_{i\downarrow}c^\dagger_{i\uparrow}|0\rangle$. Finally, the
constraint $H_{\rm con}$ is taken to be
\beq\label{con}
H_{\rm con} = s\bar{\eta}\sum_j\varphi_j^\dagger (D_j-\eta c_{j,\uparrow}c_{j,\downarrow})+h.c.
\eeq
where $\varphi_j$ is a charge $2e$ bosonic field. The constant $s$ will be
determined shortly. 
To see how this constraint removes unphysical
states that arise from the extended Hilbert space, we compute the
partition function, 
\beq\label{Z}
Z=\int [{\cal D}c\ {\cal D}c^\dagger\ {\cal D}D\ {\cal D}D^\dagger\ {\cal D}\varphi\ {\cal D}\varphi^\dagger]\exp^{-\int_0^\tau L dt},
\eeq
in Euclidean signature.
The integration over $\varphi_i$ yields a series of $\delta$-functions
which makes the integral over $D$ trivial.  The resultant Lagrangian
given by $\int d^2\eta\ \bar{\eta}\eta L_{\rm
Hubb}=\sum_{i\sigma}c_{i\sigma}^\dagger\dot
c_{i\sigma}+H_{\rm Hubb}$ is identical to that of the Hubbard model. This
constitutes the {\bf ultra-violet (UV) limit} of our theory.  As is evident,
in this limit the extended Hilbert space contracts, unphysical states
such as  $|1,0,1\rangle$, $|0,1,1\rangle$, $|1,1,1\rangle$ are set to
zero, and we identify $|1,1,0\rangle$ with $|0,0,1\rangle$.  Note there
is no contradiction between treating $D$ as fermionic and the constraint
in Eq. (\ref{con}). The constraint never governs the commutation
relation for $D$. The value of $D$ is determined by Eq. (\ref{con}) only
when $\varphi$ is integrated over. This is followed immediately by an integration over $D$ at
which point $D$ is eliminated from the theory.

The theory given above permits us to coarse
grain the system cleanly for $U\gg t$.  The energy scale associated with
$D$ is the large on-site energy $U$.  Hence, it makes sense, instead of solving the constraint, to integrate
out $D$. This will result in the {\bf low-energy (IR) theory}. 
Such an integration may be done exactly as the theory is Gaussian in $D$.
 This is not possible in previous theories. Because of the
dynamical term in the action, integration over $D$ will yield a theory
that is frequency dependent.  We identify the corresponding low-energy
theory by setting the frequency to zero.  Since the theory is Gaussian,
it suffices to complete the square in the $D$-field. To accomplish this,
we define the matrix
\beq\label{eom}
 {\cal M}_{ij}=
\left(\delta_{ij}-\frac{t}{(\omega+U)}g_{ij}\sum_\sigma c_{j,\sigma}^\dagger c_{i,\sigma}\right)
\eeq
and $b_{i}=\sum_{j}b_{ij}=\sum_{j\sigma} g_{ij}c_{j,\sigma}V_\sigma
c_{i,\bar\sigma}$. At zero frequency the Hamiltonian\cite{trlnm} is
\beq
H^{IR}_h = -t\sum_{i,j,\sigma}g_{ij}
\alpha_{ij\sigma}c^\dagger_{i,\sigma}c_{j,\sigma}+ H_{\rm int}-\frac{1}{\beta}Tr\ln{\cal M}
\nonumber
\eeq
where
\beq\label{HIR}
H_{\rm int}=-\frac{t^2}U \sum_{j,k} b^\dagger_{j}
({\cal M}^{-1})_{jk} b_{k}-\frac{s^2}U\sum_{i,j}\varphi_i^\dagger
 ({\cal M}^{-1})_{ij} \varphi_j\nonumber\\
-s\sum_j\varphi_j^\dagger c_{j,\uparrow}c_{j,\downarrow}
+\frac{st}U \sum_{i,j}\varphi^\dagger_i ({\cal M}^{-1})_{ij}
b_{j}+h.c.\;\;,
\eeq
which constitutes the true (IR) limit as the high-energy scale has
been removed. The energy scale $s$ is set by
noting that the fourth term entering our Hamiltonian can mediate spin
exchange. As the energy scale for this process is $t^2/U$, we make the
identification $s\simeq t$.  Hence, appearing at low energy is a charge $2e$
bosonic field which can either annihilate/create doubly occupied sites
or nearest-neighbour singlets. That the energy cost for double occupancy
in the IR is $t^2/U$ and not $U$ underscores the fact that the
UHB and LHB are not orthogonal. If they were, integrating out the high energy scale would not result in new charge 2$e$ degrees of freedom at low energy.  While electron number conservation is broken in
the IR, a conserved low-energy charge does exist, however\cite{cnote}:
$Q=\sum_{i\sigma}c_{i\sigma}^\dagger
c_{i\sigma}+2\sum_i\varphi^\dagger_i\varphi_i$. As Eq. (\ref{HIR}) implies, the bosons acquire dynamics
only through electron motion. Further, they lack a Fock space  of their own since all operators in the extended space have been integrated out. 
 Indeed,
on phenomenological grounds, weakly interacting bose-fermi models with a boson Fock space have been
advanced\cite{tdlee,ranninger}.  What
the current analysis lays plain is that there is a rigorous connection
between a strongly coupled bose-fermi model and the
low-energy physics of doped Mott insulators. As we will see, $\varphi_i$'s role is to provide internal structure to the electron by mediating composite excitations.

In the limit $U\to\infty$, the theory
 reduces to the restricted hopping term and the third term
in Eq. (\ref{HIR}).  Performing the $\varphi$ integration in the
partition function, we arrive at the constraint
$\delta(c_{i\uparrow}c_{i\downarrow})$.  This leads to a vanishing of
double occupancy, the correct result for $U=\infty$.  Second, for
$\varphi=0$, we have the restricted hopping term and second term in Eq.
(\ref{HIR}).  Approximating ${\cal M}_{ij}$ by its leading term, $\delta_{ij}$,
we reduce the second term to $\sum_i b_i^\dagger
b_i=\sum_{ij\ell\sigma\sigma'} g_{ij}g_{j\ell}
c^\dagger_{i,\sigma}V_\sigma
c^\dagger_{j,\bar\sigma}c_{\ell,\sigma'}V_{\sigma'}c_{j,\bar\sigma'}$ which
contains the spin-spin interaction $-(S_i\cdot S_j-n_i n_j/4)$ as well
as the three-site hopping term.  Hence, the $\varphi=0$ limit contains the
$t-J$ model, thereby establishing that the physics contained in
$\varphi_i$ is non-projective.\\
{\noindent\bf Electron Doping:}  For electron doping, the chemical
potential jumps to the bottom of the UHB and hence the degrees of
freedom that lie far away from the chemical potential no longer
correspond to double occupancy but rather double holes.  We proceed as before by extending the Hilbert space
and constructing a new Lagrangian
\beq\label{Led}
L&&=\int d^2\eta\left[\bar{\eta}\eta\sum_{i\sigma}n_{i\bar\sigma}
 c^\dagger_{i\sigma}\dot c_{i\sigma}+\sum_i \tilde D_i^\dagger\dot {\tilde D}_i-U\sum_j \tilde D_j\tilde D^\dagger_j\right.\nonumber\\
&&- t\sum_{i,j,\sigma}g_{ij}\left[c^\dagger_{i,\sigma}n_{i\bar\sigma}c_{j,\sigma}n_{j\bar\sigma}
+\tilde D_j^\dagger c^\dagger_{j,\sigma}c_{i,\sigma}\tilde D_i\right.\nonumber\\
&&-\left.\left.(\eta c_{i,\sigma}V_\sigma c_{j,\bar\sigma}\tilde D_i+h.c.)
\right]+\tilde{H}_{\rm con}\right],
\eeq
that preserves the distinct hops in the Hubbard model where the
operator $\tilde D_i$ is a fermion associated with double holes.  In this case,
the constraint is given by $\tilde{H}_{\rm
con}=s\bar{\eta}\sum_i\varphi_i(\tilde D_i-\eta c_{i\downarrow}^\dagger
c_{i\uparrow}^\dagger)+h.c.$. Two differences to note are that 1) because the
chemical potential resides in the UHB, the electron hopping term now
involves sites that are at least singly occupied and 2) the order of the
$\tilde D_i$ and $c_i$ operators is important.  If we integrate over
$\varphi_i$ and then $\tilde D_i$, all the unphysical states are removed and we
obtain as before precisely $L_{\rm Hubb}$. Hence, both
theories yield the Hubbard model in their UV limits.  They differ, however,
in the IR as can be seen by performing the integration over $\tilde D_i$. The
corresponding integral is again Gaussian and yields
\beq\label{edir}
H^{IR}_{\rm e} = -t\sum_{i,j,\sigma}
g_{ij}c^\dagger_{i,\sigma}n_{i\bar\sigma}c_{j,\sigma}n_{j\bar\sigma}+
\tilde{H}_{\rm int}-\frac{1}{\beta}Tr\ln{\cal M}\nonumber
\eeq
where
\beq\label{HIRE}
\tilde{H}_{\rm int}=-\frac{t^2}U \sum_{j,k} b^\dagger_{j}
({\cal M}^{-1})_{jk} b_{k}-\frac{s^2}U\sum_{i,j}\varphi_i
({\cal M}^{-1})_{ij} \varphi^\dagger_j\nonumber\\
-s\sum_j\varphi_j c_{j,\uparrow}c_{j,\downarrow}
-\frac{st}U \sum_{i,j}\varphi_i ({\cal M}^{-1})_{ij}
b_{j}+h.c.,
\eeq
as the IR limit of the electron-doped theory.  In addition to the hopping
term, the last term also differs from the hole-doped theory as it enters
with the opposite sign. The generalised particle-hole transformation (GPHT) that leads to the hole-doped theory is $c_{i\sigma}\rightarrow e^{i\vec Q\cdot \vec r_i}c_{i\sigma}^\dagger$ (${\vec Q}={\pi,\pi}$) augmented with $\varphi_i\rightarrow-\varphi_i^\dagger$. As $\varphi_i$ is a complex field, the GPHT interchanges the creation operators of opposite charge. We again make the identification $s\simeq t$ because the last term can also mediate spin exchange.
 When the boson vanishes, we
do recover the exact particle-hole symmetric analogue of the hole-doped
theory.  Because the field $\varphi$ now couples to double holes, the relevant creation operator has
charge $-2e$ and the conserved charge\cite{cnote} is
$\tilde{Q}=\sum_{i\sigma}c_{i\sigma}^\dagger
c_{i\sigma}-2\varphi_i^\dagger\varphi_i$.  This sign change in the
conserved charge will manifest itself as a sign change in the chemical
potential as long as $\langle\varphi_i^\dagger\varphi_i\rangle\ne 0$. Likewise, the correct $U\rightarrow\infty$ limit is obtained
as before.

To uncloak how the spectrum changes upon single-electron addition or subtraction, we derive an {\it exact} expression for the electron operator in the new low energy theory.  To this end, we translate the Lagrangian
for the hole-doped theory by a source term, that generates the canonical electron operator when the constraint is solved.  The appropriate source term that yields the canonical electron operator in the UV, is $\sum_{i\sigma} J_{i,\sigma}\left[\bar\eta\eta(1-n_{i,\bar\sigma} ) c_{i\sigma}^\dagger\nonumber + V_\sigma D_i^\dagger c_{i,\bar\sigma}\eta\right] + h.c.$. However, in the IR in which we only integrate over the heavy degree of freedom, $D_i$, the electron creation operator
\beq\label{cop}
(1-n_{i,\bar\sigma})c_{i,\sigma}^\dagger +V_\sigma \frac{t}{U} b_i^\dagger{\cal M}_{ij}^{-1} c_{j,\bar\sigma}- V_\sigma \frac{s}{U}\varphi_i^\dagger{\cal M}_{ij}^{-1}c_{j,\bar\sigma}
\eeq
contains the standard term for motion in the LHB, $(1-n_{i,\bar\sigma})c_{i,\sigma}^\dagger$ with a renormalization from spin fluctuations (second term) and a new charge $e$ excitation, $c_{i,\bar\sigma}{\cal M}_{ij}^{-1}\varphi_j^\dagger$, the IR analogue of the UHB excitation $n_{i,\bar\sigma}c_{i,\sigma}^\dagger$. Consequently, we predict that
 an electron at low energies is in a coherent superposition of the standard LHB state (modified with spin fluctuations) and a new charge $e$ state described by $c_{i,\bar\sigma}{\cal M}_{ij}^{-1}\varphi_j^\dagger$. It is the presence of these two distinct excitations that preserves the 2x sum
rule\cite{sawatzky}. 

\begin{figure}
\centering
\includegraphics[width=6.1cm,angle=-90]{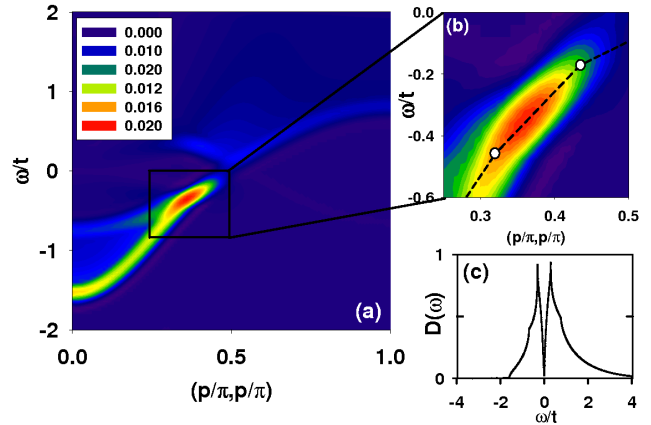}
\caption{(a) Spectral function along the nodal direction for filling $n=0.9$ and $U=10t$ at $T=0$. The lower energy branch below the chemical potential arises from a binding (which opens a pseudogap in the density of states (c)) of a hole with the charge 2e boson. (b) Two kinks occur, one at $0.15t\approx 70 meV$ and a higher-energy kink at $.5t\approx 250 meV$ at which the bifurcation obtains.}
\label{specfunc}
\end{figure}

To illustrate this physics, we offer an approximate calculation of the electron spectral function.  For the sake of the following discussion, we consider $\varphi_i$ to be spatially independent and thereby compute the electron Green function by evaluating 
$\int d\phi^* d\phi FT (\int [Dc_i^* Dc_i]  c_i(t) c_j(0) ^* 
\exp(-\int L_{\rm IR} dt))/Z$.
Further, we ignore the four-fermion term, $b_i^\dagger b_i$, as this term simply renormalizes the standard LHB band as Eq. (\ref{cop}) indicates. The key results summarised in Fig. (\ref{specfunc}) are:  1)  Below the chemical potential the electron spectral function consists of two branches. The inner curve corresponds to the standard LHB of the $t-J$ model while the outer curve the new excitation arising from the non-projective physics in the true low energy theory. The bifurcation persists for a wide range of doping ($1.3>n>0.7$) and cannot be captured by mean-field or saddle-point approximations to the Green function.  2) Bifurcation occurs at the second higher-energy kink (roughly $0.5t$ approximately 250meV for the cuprates).  3)  The difference between the two branches is largest at the momentum (0,0) and scales as $t$.  4) The presence of the two branches opens a pseudogap in the spectrum as shown in the density of states (panel c). The bifurcation, intensity of each branch and energy of the kink are in excellent agreement with recent experiment\cite{graf}. However, the bifurcation\cite{graf} has been
interpreted as evidence for spin-charge separation. Although our method is approximate, it is sufficient to capture the essence of Eq. (\ref{cop}): two distinct excitations (as seen experimentally) constitute the removal of an electron 
at low energies in a doped Mott insulator.

Whether or not $c_{i\sigma}{\cal M}_{ij}^{-1}\varphi_j^\dagger$ constitutes a true bound state
remains a conjecture at this point, though the presence of two states (Fig. (\ref{specfunc}a)) in the excitation spectrum and Eq. (\ref{cop}) are consistent with such physics. An analysis based on Bethe-Saltpeter equations is necessary.  Consequently, the physics of doped Mott insulators turns on precisely the same kinds
of problems that arise in other instances of strong coupling such as nuclear structure and confinement.

\acknowledgements We thank Tudor Stanescu for comments, and the NSF DMR-0605769 for partial support.


\begin{thebibliography}{99}
\bibitem{trlnm}The leading term of $k_BT$Tr$\ln{\cal M}$ scales as $k_BT(t/U)^2$ and hence vanishes at $T=0$. In all calculations, we replaced ${\cal M}^{-1}_{ij}$ with its leading term in $t/U$, namely $\delta_{ij}$.
\bibitem{cnote}In this expression for $Q$, $\varphi^\dagger$ should be interpreted as a creation operator with the appropriate charge.
\end{thebibliography}
\end{document}